\documentclass[aps,prb,twocolumn,superscriptaddress,showpacs]{revtex4}
\usepackage{amssymb}

\usepackage{graphicx}
\usepackage{dcolumn}
\usepackage{bm}

\begin{document}

\title{Ferroelectricity of structural origin in spin-chain compounds Ca$_3$Co$_{2-x}$Mn$_x$O$_6$}

\author{J. Shi}
\affiliation{Department of Physics, University of Science and Technology of China, Hefei, Anhui 230026, People's Republic of China}

\author{J. D. Song}
\affiliation{Hefei National Laboratory for Physical Sciences at Microscale, University of Science and Technology of China, Hefei, Anhui 230026, People's Republic of China}

\author{J. C. Wu}
\affiliation{Hefei National Laboratory for Physical Sciences at Microscale, University of Science and Technology of China, Hefei, Anhui 230026, People's Republic of China}

\author{X. Rao}
\affiliation{Hefei National Laboratory for Physical Sciences at Microscale, University of Science and Technology of China, Hefei, Anhui 230026, People's Republic of China}

\author{H. L. Che}
\affiliation{Department of Physics, University of Science and Technology of China, Hefei, Anhui 230026, People's Republic of China}

\author{Z. Y. Zhao}
\affiliation{Hefei National Laboratory for Physical Sciences at Microscale, University of Science and Technology of China, Hefei, Anhui 230026, People's Republic of China}
\affiliation{Fujian Institute of Research on the Structure of Matter, Chinese Academy of Sciences, Fuzhou, Fujian 350002, People's Republic of China}

\author{H. D. Zhou}
\affiliation{Department of Physics and Astronomy, University of Tennessee, Knoxville, Tennessee 37996, USA}

\affiliation{National High Magnetic Field Laboratory, Florida State University, Tallahassee, Florida 32306-4005, USA}

\author{J. Ma}
\affiliation{Department of Physics and Astronomy, University of Tennessee, Knoxville, Tennessee 37996, USA}

\affiliation{Collaborative Innovation Center of Advanced Microstructures, Nanjing, Jiangsu 210093, People's Republic of China}

\affiliation{Key Laboratory of Artificial Structures and Quantum Control, Department of Physics and Astronomy, Shanghai Jiao Tong University, Shanghai 200240, People's Republic of China}

\author{R. R. Zhang}
\affiliation{High Magnetic Field Laboratory, Chinese Academy of Sciences, Hefei, Anhui 230031, People¡¯s Republic of China}

\author{L. Zhang}
\affiliation{High Magnetic Field Laboratory, Chinese Academy of Sciences, Hefei, Anhui 230031, People¡¯s Republic of China}

\author{X. G. Liu}
\affiliation{Hefei National Laboratory for Physical Sciences at Microscale, University of Science and Technology of China, Hefei, Anhui 230026, People's Republic of China}

\author{X. Zhao}
\affiliation{School of Physical Sciences, University of Science and Technology of China, Hefei, Anhui 230026, People's Republic of China}

\author{X. F. Sun}
\email{xfsun@ustc.edu.cn}
\affiliation{Hefei National Laboratory for Physical Sciences at Microscale, University of Science and Technology of China, Hefei, Anhui 230026, People's Republic of China}

\affiliation{Collaborative Innovation Center of Advanced Microstructures, Nanjing, Jiangsu 210093, People's Republic of China}

\affiliation{Key Laboratory of Strongly-Coupled Quantum Matter Physics, Chinese Academy of Sciences, Hefei, Anhui 230026, People's Republic of China}

\date{\today}

\begin{abstract}

We report a systematic study of the structure, electric and magnetic properties of Ca$_3$Co$_{2-x}$Mn$_x$O$_6$ single crystals with $x =$ 0.72 and 0.26. The DC and AC magnetic susceptibilities display anomalies with characteristic of the spin freezing. The crystals show ferroelectric transition at 40 K and 35 K ($T_{FE}$) for $x =$ 0.72 and 0.26, respectively, with a large value of 1400 $\mu$C/m$^2$ at 8 K for electric polarization ($P_c$) along the spin-chain ($c$-axis) direction. Interestingly, the electric polarization perpendicular to the chain direction ($P_{ab}$) can also be detected and has value of 450 and 500 $\mu$C/m$^2$ at 8 K for the $x =$ 0.72 and 0.26 samples, respectively. The specific heat and magnetic susceptibility show no anomaly around $T_{FE}$, which means that the electric polarization of these samples has no direct relationship with the magnetism. The X-ray diffraction and the Raman spectroscopy indicate that these samples may undergo Jahn-Teller distortions that could be the reason of electric polarization.

\end{abstract}



\maketitle

\section{Introduction}

Multiferroic materials have attracted considerable attention in recent years due to their potential applications in magnetoelectronics, spintronics and magnetic memory technology.\cite{Multiferroics1, Multiferroics2, LiCuO, CoCrO, NiVO, CaCoMnO} Ca$_3$Co$_{2-x}$Mn$_x$O$_6$ ($x \approx 1$) was found to be a typical system with magnetism-driven ferroelectricity.\cite{CaCoMnO, ab inito, Magnetization, high-field measures, ferroelectricity modulation} It is composed of the $c$-axis spin chains consisting of magnetic ions with alternating face-sharing CoO$_6$ trigonal prisms and MnO$_6$ octahedra, as shown in Fig. 1(a). Both the Co$^{2+}$ ions and Mn$^{4+}$ ions are in the high-spin states.\cite{ab inito} Because of the much stronger intrachain interaction than the interchain one and the strong Ising-like anisotropy, it can be characterized by a one-dimensional Ising model.\cite{ab inito} A special up-up-down-down antiferromagnetic structure was found, caused by the frustrated ferromagnetic nearest-neighbor and antiferromagnetic next-nearest-neighbor interactions within the spin chain.\cite{CaCoMnO} The alternating Co$^{2+}$ and Mn$^{4+}$ ionic order breaks the inversion symmetry and induces an electric polarization along the chain direction at 16.5 K.\cite{CaCoMnO} The electric polarization is destroyed when applying magnetic field along the $c$ axis, which was proposed to be due to the magnetic-field-induced transition from the up-up-down-down state to the up-up-up-down spin-solid state.\cite{Magnetization} Whereas, a different mechanism for the suppression of electric polarization with magnetic fields near 10 T was proposed to be flopping of the Mn$^{4+}$ spins into the ${ab}$ plane.\cite{high-field measures} Moreover, a slight destruction of the Co/Mn ionic order could significantly enhance the ferroelectricity. For example, it was found that the polycrystal sample of Ca$_3$Co$_{1.08}$Mn$_{0.92}$O$_6$ has a ferroelectric transition at $T_{FE}$ $\sim$ 30 K.\cite{ferroelectricity modulation} It was explained that the competition between the ionic order and spin frustration is crucial for improving the ferroelectric performance.\cite{ferroelectricity modulation} It would be interesting to study the electric and magnetic properties of Ca$_3$Co$_{2-x}$Mn$_{x}$O$_6$ with even larger deviation of the composition from $x \approx 1$. Some earlier studies were mainly performed with polycrystal samples and focused on their magnetism. In particular, an amazing memory effect was found in Ca$_3$Co$_{1.62}$Mn$_{0.38}$O$_6$ single crystal.\cite{Ca$_3$Co$_{2-x}$Mn$_{x}$O$_6$, Ca$_3$Co$_{1.62}$Mn$_{0.38}$O$_6$, memory effect} However, the study about their electric properties are still missing.

In this work, we grew single crystals of Ca$_3$Co$_{2-x}$Mn$_x$O$_6$ with $x =$ 0.72 and 0.26, and studied the structure, magnetism, thermodynamic property, ferroelectricity, and dielectric constant. It was found that along the spin-chain direction both the magnitude of electric polarization and transition temperature are enhanced several times, compared with those of Ca$_3$CoMnO$_6$.\cite{CaCoMnO, Magnetization} In addition, the electric polarization perpendicular to the chain direction was also detected. This is the first time in the experiment observing the polarization perpendicular to the $c$ direction in Ca$_3$Co$_{2-x}$Mn$_x$O$_6$ system. The origin of the ferroelectricity in the $x =$ 0.72 and 0.26 samples is found to be likely the lattice distortion instead of magnetism.

\begin{figure}
\includegraphics[clip,width=8.5cm]{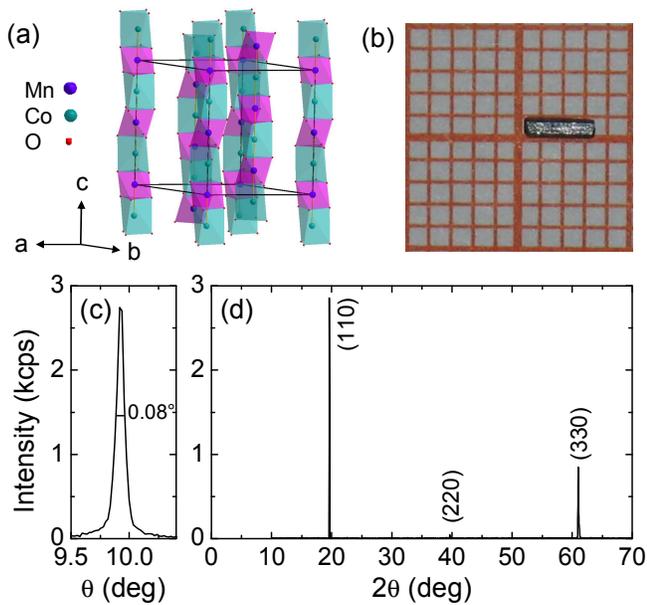}
\caption{(color online) (a) Crystal structure of Ca$_3$CoMnO$_6$. The Ca ions are omitted for clarity. (b) The photo of a selected Ca$_3$Co$_{1.28}$Mn$_{0.72}$O$_6$ single crystal, the grid scale is 1 mm. (c) X-ray rocking curve of the (110) reflection, with a FWHM of only 0.08$^{\circ}$, for an $x =$ 0.72 single crystal. (d) The ($hh$0) diffraction pattern of this sample.}
\end{figure}

\section{Experiments}

Single crystals of Ca$_3$Co$_{2-x}$Mn$_x$O$_6$ ($x =$ 0.72 and 0.26) were grown using a K$_2$CO$_3$ flux.\cite{Crystal growth} In the first step, single-phase Ca$_3$Co$_{4-2x}$Mn$_{2x}$O$_9$ powder was obtained by calcining a mixture of CaCO$_3$, MnO$_2$ and CoO in air at 950 $^{\circ}$C for 3 days with several intermediate grindings. Mixture of Ca$_3$Co$_{4-2x}$Mn$_{2x}$O$_9$ and K$_2$CO$_3$ in a weight ratio of 1 : 7 was loaded into aluminum crucibles and soaked at 950 $^{\circ}$C for 6 h and then slowly cooled at a rate of 0.25 $^{\circ}$C/h to 880 $^{\circ}$C, before finally cooled down to room temperature at rate of 100 $^{\circ}$C/h. This method yielded long-bar shaped single crystals with the longest dimension along the $c$ axis, as shown in Fig. 1(b). The maximum size of single crystals is about 5 mm ${\times}$ 1.5 mm ${\times}$ 0.6 mm.

The samples were characterized by X-ray diffraction (XRD) at room temperature and the chemical composition was checked via X-ray fluorescence spectrometer (XRF). The low-temperature XRD was also carried out to investigate the structural phase transition. The Raman scattering measurements were performed by using a Horiba Jobin Yvon T64000 Micro-Raman instrument with a Torus laser ($\lambda$ = 532 nm) as an excitation source in a backscattering geometry, and the laser power was kept at $\sim$ 1 mW to avoid local heating effect. The DC and AC magnetic susceptibilities were measured by using a superconducting quantum interference device-vibrating sample magnetometer (SQUID-VSM, Quantum Design). The specific heat was measured by a relaxation method using a commercial physical property measurement system (PPMS, Quantum Design). For electric polarization and dielectric constant measurements, the crystals were cut into thin-plate shape and polished, and then annealed at 600 $^{\circ}$C for 6 hours to remove the strain caused by polishing. The samples were poled in an electric field $E \approx$ 10 kV/cm from high temperature ($>$ 50 K) to 8 K and then $P_c$ and $P_{ab}$ as a function of temperature were obtained by integrating the pyroelectric current measured by an electrometer (model 6517B, Keithley). The dielectric constant along the $c$ axis was measured by using HP4294 impedance analyzer.

\section{Results and Discussion}

The XRD patterns at room temperature indicate that both the $x =$ 0.72 and 0.26 samples have a K$_4$CdCl$_6$-type crystal structure (space group $R\overline{3}c$), the same as that of the $x =$ 1 sample.\cite{lattice structure} Figure 1(c) shows the X-ray rocking curve of the (110) diffraction for an $x =$ 0.72 single crystal. The peak is very narrow with the full width at half maximum (FWHM) of 0.08$^\circ$, indicating the good crystallization of the crystal. The X-ray diffraction pattern of the ($hh$0) plane is shown in Fig. 1(d). These XRD results indicate that the single crystals are of pure phase and good crystallinity.

\begin{figure}
\includegraphics[clip,width=6cm]{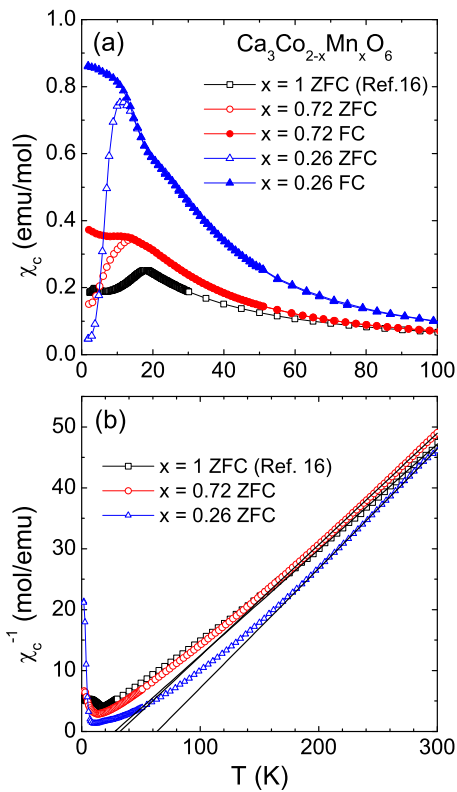}
\caption{(color online) (a) Magnetic susceptibility ${\chi}_c$ as a function of $T$ for $x$ = 0.72 and 0.26 single crystals in 1000 Oe field along the $c$ axis. (b) Temperature dependence of the inverse susceptibility ${\chi}_c^{-1}$. The solid lines are the Curie-Weiss fitting to the high-$T$ data. For comparison, data of the $x =$ 1 sample taken from Ref. \onlinecite{Flint} are also shown.}
\end{figure}

Figure 2(a) shows the magnetic susceptibility ${\chi}_c$ as a function of temperature under zero-field cooling (ZFC) and field cooling (FC) conditions for the $x =$ 0.26 and 0.72 single crystals, in comparison with ${\chi}_c$ of $x =$ 1 single crystal taken from an earlier literature.\cite{Flint} The applied magnetic field is $H =$ 1000 Oe and along the $c$ axis. As shown in Fig. 2(a), ${\chi}_c(T)$ of $x =$ 1 shows a broad peak around 17 K, which is caused by the onset of up-up-down-down magnetic order.\cite{CaCoMnO} However, the experimental results of AC magnetic susceptibility and neutron diffraction revealed that only finite-size magnetic domains develop at the magnetic transition and these domains exhibit an additional spin freezing at a lower temperature.\cite{CaCoMnO} Thus, the magnetic ground state of $x =$ 1 is a co-existence of magnetic order and spin freezing. This kind of magnetic state can be formed with smaller $x$, which for example was confirmed in the $x =$ 0.75 polycrystal sample by AC magnetic susceptibility and neutron powder diffraction.\cite{Rayaprol, Order by static disorder} Apparently, our $x =$ 0.72 sample exhibits similar magnetic properties to the $x =$ 1 and 0.75 cases; the broad peak of ${\chi}_c(T)$ at $T_{max} =$ 14 K and the separation of the ZFC and FC curves below $T_{max}$ should be attributed to magnetic ordering and spin freezing, respectively. For the $x =$ 0.26 samples, there are two clear anomalies in the ${\chi}_c(T)$ curves. One is the broad peak at $T_{max} =$ 11 K, below which the FC and ZFC curves separate with each other. Another one is a clear upturn of ${\chi}_c(T)$ at $T <$ 18 K. Both of these two features in ${\chi}_c(T)$ of $x =$ 0.26 agree well with the case of $x =$ 0 and 0.20 polycrystal samples.\cite{Ca$_3$Co$_{2-x}$Mn$_{x}$O$_6$} Thus, the 18 K upturn should be caused by the intrachain ferromagnetic order, and the ZFC-FC curves separation is attributed to a spin freezing. Therefore, the $x =$ 1, 0.72 and 0.26 samples all appear a spin freezing, which is likely a common feature in Ca$_3$Co$_{2-x}$Mn$_x$O$_6$, and is probably due to chemical disorder and competing magnetic interactions.\cite{Ca$_3$Co$_{1.62}$Mn$_{0.38}$O$_6$}

Figure 2(b) shows the inverse magnetic susceptibility ${\chi}_c^{-1}$. The high-temperature data can be fitted with the Curie-Weiss law ($\chi = \frac{C}{T - \theta_p}$), leading to the Weiss temperatures $\theta_p =$ 28.0, 37.4 and 63.5 K, and the effective moments $\mu$$_{eff} =$ 6.79, 6.50 and 6.38 $\mu_{B}$/f.u. for the $x =$ 1, 0.72 and 0.26 samples, respectively.

\begin{figure}
\includegraphics[clip,width=8.5cm]{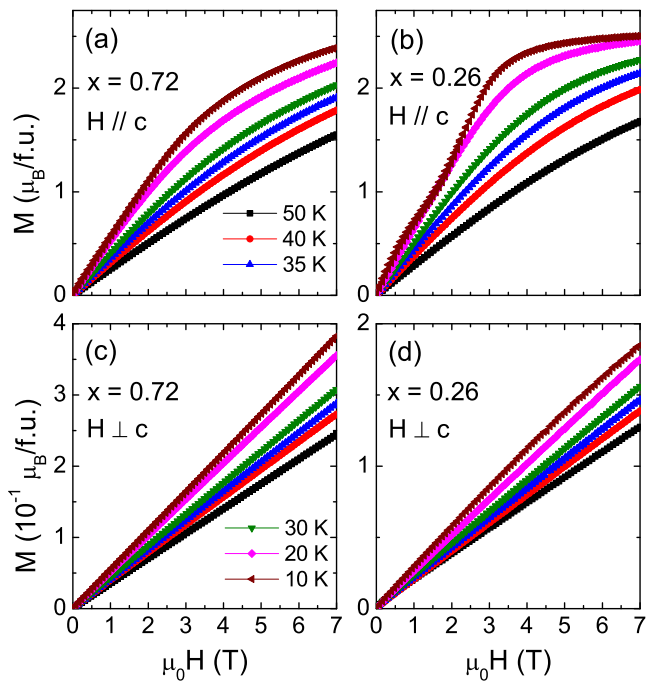}
\caption{(color online) The magnetization curves with magnetic field parallel and perpendicular to the $c$ direction for $x =$ 0.72 and 0.26 single crystals.}
\end{figure}

Figure 3 shows the $M(H)$ curves of the $x =$ 0.72 and 0.26 single crystals at 50 to 10 K with different field directions. The behaviors are comparable to those of the polycrystal samples and can be understood with the Ising-like Co spins and quasi-isotropic Mn spins.\cite{high-field measures}. For $H \parallel c$, the $M(H)$ curves at 50 K are nearly linear for both $x =$ 0.72 and 0.26 samples. With decreasing temperature, the $M(H)$ curves show clear curvatures, similar to the case of $x =$ 0.75 and 0.20 polycrystal samples.\cite{Ca$_3$Co$_{1.25}$Mn$_{0.75}$O$_6$, Ca$_3$Co$_{2-x}$Mn$_{x}$O$_6$} It is notable that the 10 K $M(H)$ curve of the $x =$ 0.26 sample displays a feature of magnetization step, which is similar to the magnetization of $x =$ 0.\cite{Crystal growth} All these $c$-axis magnetization behaviors are mainly dominated by the the Ising-like Co spins.\cite{high-field measures} For $H \perp c$, all the $M(H)$ curves show a linear increase with increasing field and the magnitudes are much smaller than the case of $H \parallel c$. The magnetization for $H \perp c$ should be mainly contributed by the quasi-isotropic Mn spins, and the difference between the the $x =$ 0.72 and 0.26 samples is supportive for this.

\begin{figure}
\includegraphics[clip,width=8.5cm]{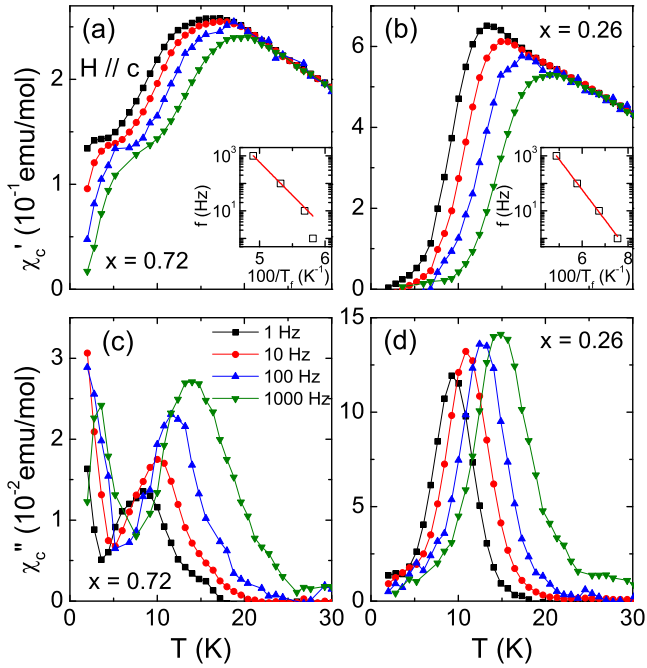}
\caption{(color online) Real and imaginary parts ($\chi'_c$ and $\chi''_c$) of the AC susceptibility of $x =$ 0.72 and 0.26 single crystals.  The AC magnetic field is 3.8 Oe and the frequencies are 1, 10, 100, and 1000 Hz. The insets show frequency vs inverse temperature of the maximum of ${\chi}'$. The red lines are fitting results with the Arrhenius behavior.}
\end{figure}

To further confirm the occurrence of spin freezing in $x =$ 0.72 and 0.26 samples, the AC magnetic susceptibility were measured at low temperatures with various frequencies. As shown in Fig. 4, large peaks are clearly observed in both the real and imaginary parts of the AC susceptibility. The peaks exhibit strong frequency dependence, shifting to higher temperature with increasing frequency. The Arrhenius behavior, $f = f_0$exp$[-E_a/k_BT_f]$, is used to the quantitative analysis of the shift, where $T_f$ is the temperature of the maximum ${\chi}'$.\cite{CaCoMnO, Spin glass} The frequency shift of the $x =$ 0.26 sample is well fitted with $E_a/k_B =$ 253 K and $f_0 =$ 230 MHz, as shown in the inset to Fig. 4(b). Another coefficient $K = \Delta T_f/T_f\Delta (log f)$ can also be used to characterize the spin glass.\cite{Spin glass} The $K$ of the $x =$ 0.26 data is determined to be 6 ${\times}$ 10$^{-2}$, within the range 5 ${\times}$ 10$^{-3}$ -- 8 ${\times}$ 10$^{-2}$ for spin glass.\cite{Spin glass} All these indicate that the $x =$ 0.26 sample has a spin-glass phase at low temperatures. For the $x =$ 0.72 sample, the peaks in AC magnetic susceptibility also exhibit strong frequency dependence. However, the $T_f$ exhibits a discrepancy from the Arrhenius behavior, as shown in the inset to Fig. 4(a). The similar behavior was found in the $x =$ 0.96 sample, which is due to the co-existence of magnetic order.\cite{CaCoMnO}

\begin{figure}
\includegraphics[clip,width=6.5cm]{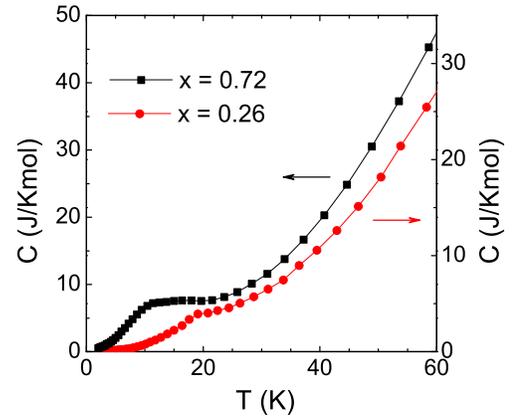}
\caption{(color online) Specific heat of the $x$ = 0.72 and 0.26 single crystals.}
\end{figure}

Figure 5 shows the specific heat of the $x =$ 0.72 and 0.26 single crystals at $T <$ 60 K. The specific heat of the $x =$ 0.72 sample shows a broad anomaly from 10 to 20 K, similar to that in the $x =$ 0.75 polycrystal sample,\cite{Ca$_3$Co$_{1.25}$Mn$_{0.75}$O$_6$} which can be attributed to the magnetic order. For the $x =$ 0.26 sample, the specific heat shows a weak anomaly around 18 K, which is associated with the fast increase in the magnetic susceptibility, and can be attributed to the intrachain ferromagnetic order.

\begin{figure}
\includegraphics[clip,width=8.5cm]{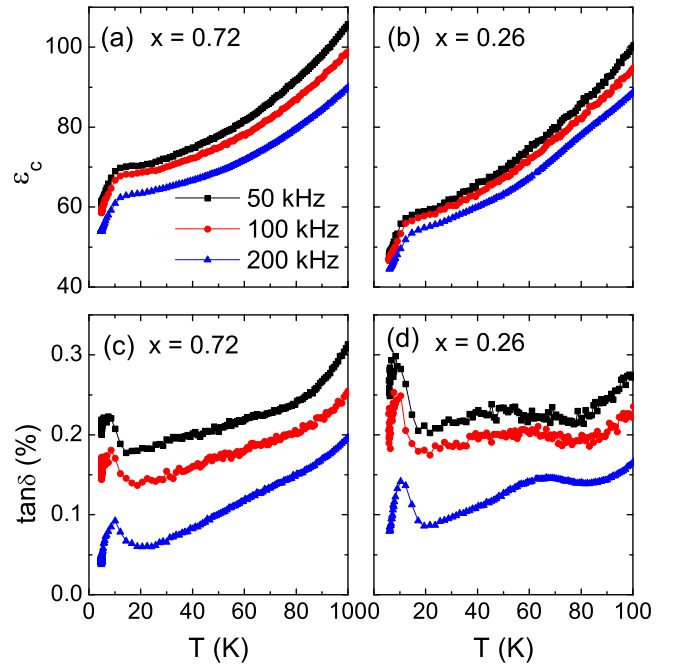}
\caption{(color online) Real part of dielectric constant and loss factor of the $x =$ 0.72 and 0.26 single crystals. The frequencies are 50, 100, and 200 kHz.}
\end{figure}

Figure 6 shows the temperature-dependent real part of dielectric constant and loss factor of the $x =$ 0.72 and 0.26 single crystals. Both of them show a clear frequency dependence. The small value of $tan\delta$ indicates that these samples have very high insulation, and no space charge contributions below 100 K.\cite{Ca$_3$Co$_2$O$_6$} Another feature is that both the real part of dielectric constant and loss factor show anomalies around $T_{max}$. The real part of dielectric constant present an abrupt decrease but the loss factor shows a peak. In this regard, the dielectric properties of these samples seem to have some relationship with the spin freezing.

\begin{figure}
\includegraphics[clip,width=8.5cm]{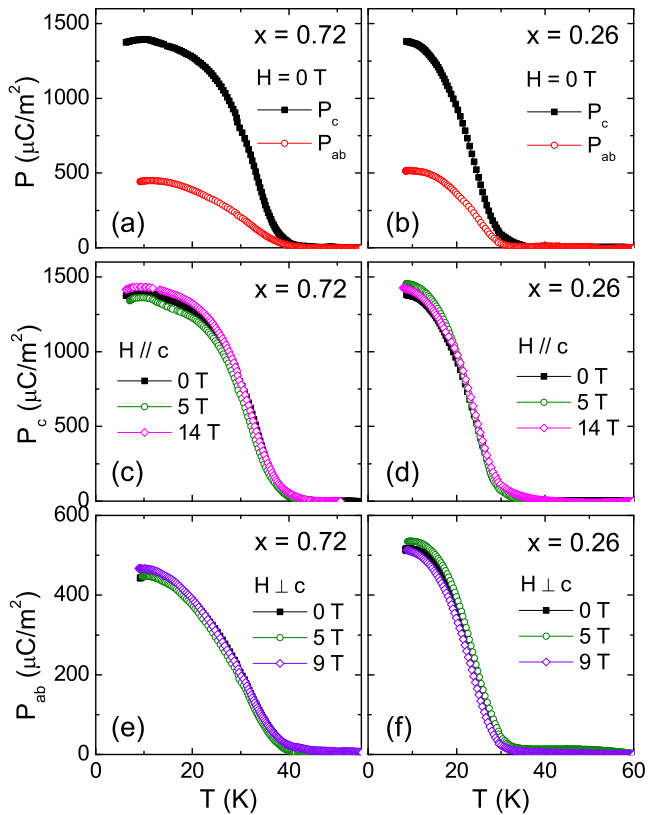}
\caption{(color online) (a,b) Electric polarization $P_ c$ and $P_{ab}$ as a function of temperature for the $x =$ 0.72 and 0.26 single crystals. (c,d) $P_c(T)$ data in different magnetic fields. (e,f) $P_{ab}(T)$ data in different magnetic fields.}
\end{figure}

It is known that ferroelectricity appears in Ca$_3$CoMnO$_6$, which was believed to be related to the special up-up-down-down spin structure with the alternating Co$^{2+}$ and Mn$^{4+}$ ionic order.\cite{CaCoMnO, Magnetization} The transition temperature, below which electric polarization shows up, is about 16.5 K and the polarization parallel to the $c$ axis ($P_c$) is about 310 $\mu$C/m$^2$ at 2 K.\cite{Magnetization} Note that there is no electric polarization perpendicular to the $c$ axis ($P_{ab}$).\cite{Magnetization} However, our $x =$ 0.72 and 0.26 single crystals display very different electric polarization behaviors. First, both $P_c$ and $P_{ab}$ can be detected, as shown in Figs. 7(a) and 5(b). This is the first time to observe the polarization perpendicular to the $c$ direction in Ca$_3$Co$_{2-x}$Mn$_x$O$_6$ system. Second, the electric polarization transition temperatures ($T_{FE}$) are 40 and 35 K for the $x =$ 0.72 and 0.26 single crystals, respectively, much higher than that of Ca$_3$CoMnO$_6$. Third, the magnitude of polarization is much larger in the $x =$ 0.72 and 0.26 single crystals. The $P_c$ gradually increases with lowering temperature and reaches 1400 $\mu$C/m$^2$ at 8 K for two samples, while the $P_{ab}$ are 450 and 500 $\mu$C/m$^2$ at 8 K for $x =$ 0.72 and 0.26, respectively. Since the $P_{ab}$ are nearly one third of the $P_c$ in both samples, the possibility of $P_{ab}$ caused by a small angle deviation from the $c$ direction can be ruled out. In passing, it should be noted that no electric polarization could be observed if the measurements were carried out without poling.

Figures 7(c--f) show the $P_{ab}(T)$ and $P_{c}(T)$ data for $x =$ 0.72 and 0.26 single crystals in different magnetic fields. Here, the magnetic field was applied before cooling, and the measurement with magnetic filed applied after cooling gave the same results. Apparently, even high magnetic fields can hardly change either the transition temperature or the magnitude of $P$. In addition, the DC and AC magnetic susceptibilities, specific heat and dielectric constant do not show any anomaly around $T_{FE}$. These results indicate that the electric polarization appearing in the $x =$ 0.72 and 0.26 samples has no direct relationship with the magnetism, in contrast to the magnetism-driven ferroelectricity in the $x =$ 1 material. One may argue that the observed $P$ is caused by some extrinsic effects, in particular, the space charges trapped at the grain boundaries or possible defects as reported elsewhere.\cite{space charges} However, this possibility can be ruled out because the electric polarizations of these samples can be well reproduced on many different pieces of single crystals. Therefore, the electric polarization appearing at rather high temperatures may originate from the lattice distortion like the case of some $AB$O$_3$ perovskites, such as LuCrO$_3$ and SmCrO$_3$ with magnetic Cr$^{3+}$ ions at the $B$ site.\cite{LuCrO$_3$, SmCrO$_3$} In these materials, the neutron pair distribution function illustrated that the electric polarization could be originated from the Jahn-Teller distortions of Cr$^{3+}$ octahedra, whereas the macroscopic crystal structure was unchanged.\cite{YCrO$_3$, SmCrO$_3$_1, LaMn$_3$Cr$_4$O$_12$} It is worth of mentioning that an earlier theory has proposed a kind of Jahn-Teller-distortion-induced ferroelectric mechanism in Ca$_3$CoMnO$_6$,\cite{Jahn-Teller distortion} which agrees very well with the experimental results of electric polarization for $x =$ 0.72 and 0.26 single crystals. In detail, it was reported that in Ca$_3$CoMnO$_6$ the Co$^{2+}$($d^7$) ions at the trigonal sites are in high-spin state ($S =$ 3/2) with unevenly filled degenerate $d$ states and a large orbital magnetic moment of 1.7 $\mu_B$, and hence have Jahn-Teller instability. In view of this, if CoO$_6$ trigonal prisms are theoretically assumed to undergo Jahn-Teller distortion, it will of course affect the electron configuration of Co$^{2+}$ ions because of the change of crystal-electric-field environment, leading to a re-distribution of orbital moment on Co$^{2+}$. Based on this, two kinds of optimized structures resulting from different types of Jahn-Teller distortions were obtained by LDA+$U$+SOC calculations in terms of the magnitude of the orbital moment on Co$^{2+}$. Therein, an optimized structure with low orbital moment on Co$^{2+}$ (i.e., 0.56 $\mu_B$) removes $C_3$ rotational symmetry leading to appearance of nonzero electric polarization both along and perpendicular to the $c$ direction.\cite{Jahn-Teller distortion} Based on these theoretical results, it is most likely that the electric polarization of $x =$ 0.72 and 0.26 are induced by Jahn-Teller distortions. To check this kind of electric-polarization mechanism, we further probed the low-temperature crystal structure of $x =$ 0.72 and 0.26 samples by using XRD and Raman Spectroscopy techniques.

\begin{figure}
\includegraphics[clip,width=7cm]{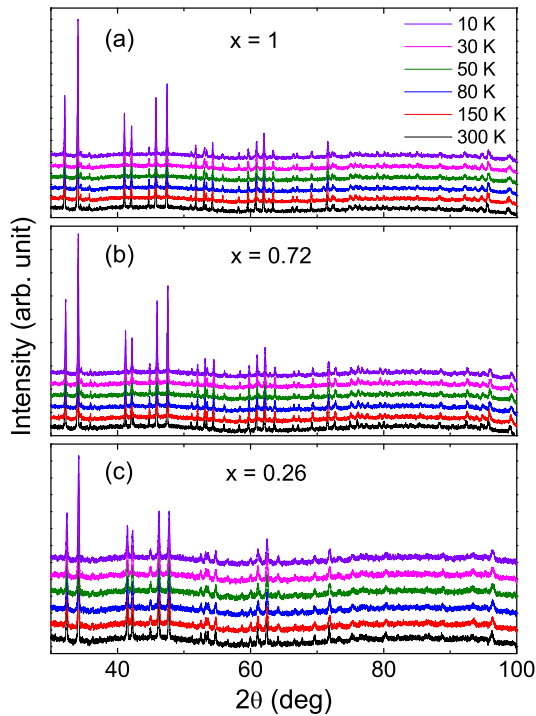}
\caption{(color online) XRD patterns at different temperatures for Ca$_3$Co$_{2-x}$Mn$_{x}$O$_6$ ($x =$ 1, 0.72 and 0.26) polycrystal samples.}
\end{figure}

The XRD of Ca$_3$Co$_{2-x}$Mn$_x$O$_6$ ($x =$ 1, 0.72 and 0.26) polycrystal samples were measured at 10--300 K, as shown in Fig. 8. With decreasing temperature, no extra diffraction peaks appear, indicating no structural phase transition. The structures of the $x =$ 1, 0.72 and 0.26 samples are rhombohedral with the space group $R\overline{3}c$ at all temperatures.

\begin{figure}
\includegraphics[clip,width=8cm]{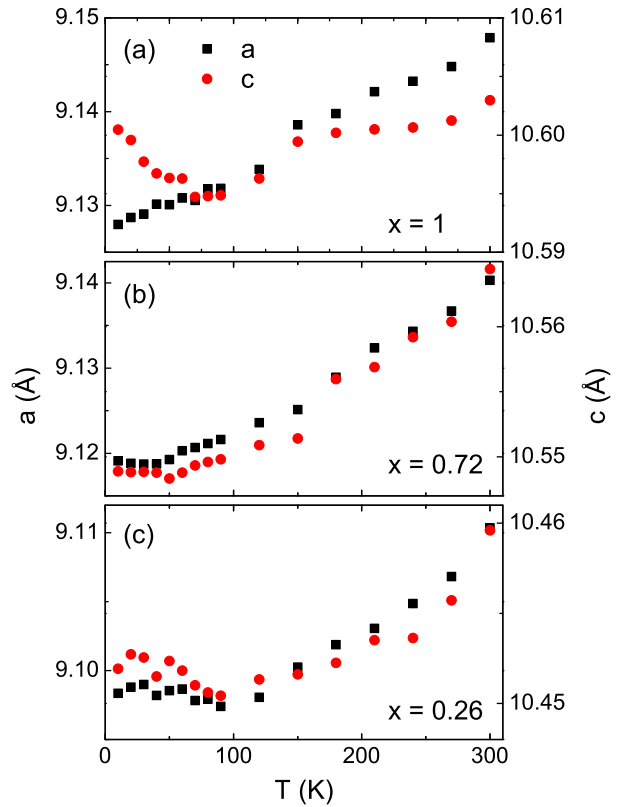}
\caption{(color online) Lattice constants as a function of temperature for Ca$_3$Co$_{2-x}$Mn$_x$O$_6$ ($x =$ 1, 0.72 and 0.26) samples.}
\end{figure}

\begin{figure}
\includegraphics[clip,width=6.5cm]{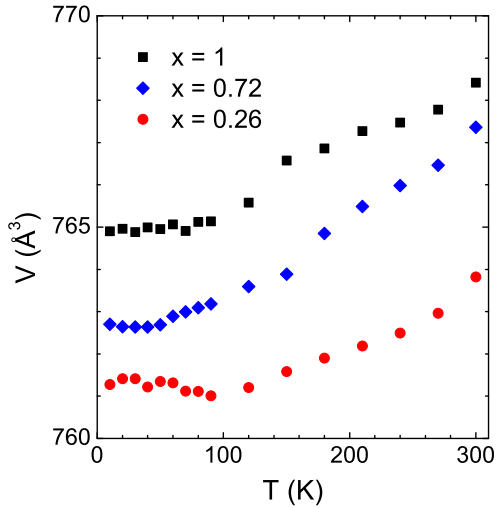}
\caption{(color online) Temperature dependence of the volume of the unit cell for Ca$_3$Co$_{2-x}$Mn$_x$O$_6$ ($x$ = 1, 0.72 and 0.26) samples. The data of the $x =$ 0.72 and 0.26 samples are shifted upward by 3 and 12 \AA$^3$, respectively.}
\end{figure}

The temperature dependence of lattice constants are shown in Fig. 9. Apparently, anomalies can be observed in all samples. For $x =$ 1, the lattice constant $a$ decreases almost linearly with decreasing temperature, and is in good agreement with the principle of thermal expansion. However, the lattice constant $c$ changes in a complicated way with temperature. It decreases with decreasing temperature until 70 K, while suddenly increases below 70 K. It is an abnormal phenomenon, which has never been reported in Ca$_3$CoMnO$_6$. Note that at $T <$ 70 K the lattice is elongated only in the $c$ direction and keeps the threefold rotational symmetry. The lattice constants $a$ and $c$ as a function of $T$ for the $x =$ 0.72 sample display an anomaly at $\sim$ 40 K. The $a$ and $c$ decrease monotonically with decreasing temperature and become nearly constants at $T \le$ 40 K, indicating that the lattice has some anomalies below 40 K. This temperature is the same as the $T_{FE}$. The $a$ and $c$ of the $x =$ 0.26 sample not only exhibit similar anomaly to that of the $x =$ 0.72 but also have anomaly at about 90 K, below which the $a$ and $c$ increase with decreasing temperature like the $c$ of $x =$ 1 sample. However, this temperature is different from the $T_{FE}$ of $x =$ 0.26 sample. The anomalies here may have some relationship with Ca$_3$Co$_2$O$_6$ and it will be discussed below. The temperature dependence of the volume of the unit cell are displayed in Fig. 10. Again, it is notable that the cell volumes $V(T)$ display anomalies for all samples.

\begin{figure}
\includegraphics[clip,width=7.5cm]{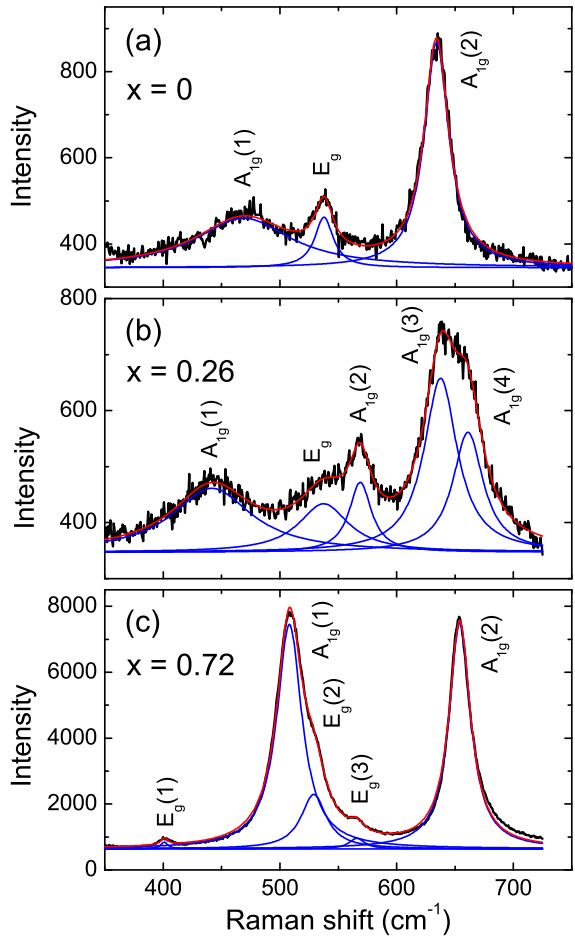}
\caption{(color online) Room-temperature Raman spectra of Ca$_3$Co$_{2-x}$Mn$_x$O$_6$ ($x =$ 0, 0.26 and 0.72). The red lines are fitting results using the Lorentzian equation and the blue lines are the decomposed fitting curves of the individual Raman-active modes. }
\end{figure}

\begin{figure}
\includegraphics[clip,width=7.5cm]{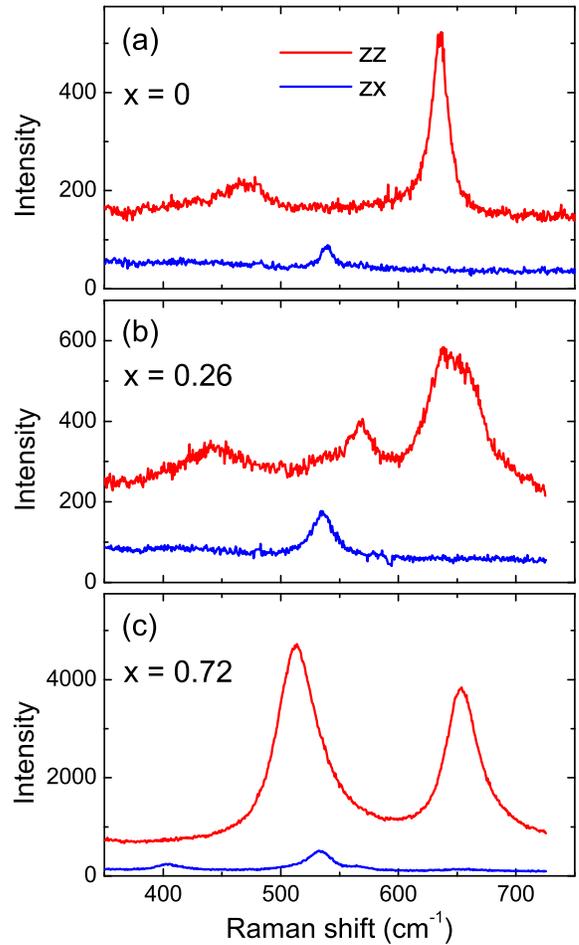}
\caption{(color online) Polarized Raman spectra of Ca$_3$Co$_{2-x}$Mn$_x$O$_6$ ($x =$ 0, 0.26 and 0.72) with scattering configuration of $zz$ and $zx$ at room temperature.}
\end{figure}

Figure 11 shows the Raman spectra of Ca$_3$Co$_{2-x}$Mn$_x$O$_6$ single crystals with $x =$ 0, 0.26, and 0.72, which were measured at room temperature in the frequency range 350--725 cm$^{-1}$. The direction of the polarization of the incident laser is along the $c$ axis. The data are fitted with Lorentzian function and decomposed into individual components, as shown in Fig. 11. The selection rule predicts 24 Raman-active modes (4 $A_{1g}$ and 20 $E_g$), 28 infrared-active modes (6 $A_{2u}$ and 22 $E_u$), 11 inactive modes (5 $A_{1u}$ and 6$A_{2g}$), and 3 acoustic translational modes ($A_{2u}$ and 2 $E_u$).\cite{Raman} In order to determine the Raman modes, polarized Raman spectra of Ca$_3$Co$_{2-x}$Mn$_x$O$_6$ single crystals in two different scattering configurations ($zz$ and $zx$) were also done and shown in Fig. 12. The $A_{1g}$ mode and the $E_g$ mode can be observed in $zz$ and $zx$ scattering configuration, respectively. The $E_{g}$ modes are related to the vibration of Cobalt ions in the octahedra, Calcium and Oxygen ions. However, the $A_{1g}$ modes are mainly attributed to the vibration of Oxygen ions. Base on this result, the individual Lorentzian components in Fig. 11 are characterized. In the case of $x =$ 0, three peaks were observed at 468, 537 and 633 cm$^{-1}$, as shown in Fig. 11(a), which are corresponding to the $A_{1g}$(1), $E_g$ and $A_{1g}$(2) modes, respectively. In the case of $x =$ 0.26, other two peaks at 568 and 661 cm$^{-1}$ are detected besides the same three modes as the $x =$ 0 sample, as shown in Fig. 11(b). These are $A_{1g}$ modes. In the $x =$ 0.72 sample, the Raman spectrum changes in such a way that the main features observed in the $x =$ 0 sample are no longer detected. Two stronger peaks at 507 and 654 cm$^{-1}$ are corresponding to $A_{1g}$ modes, as shown in Fig. 11(c).

\begin{figure}
\includegraphics[clip,width=7cm]{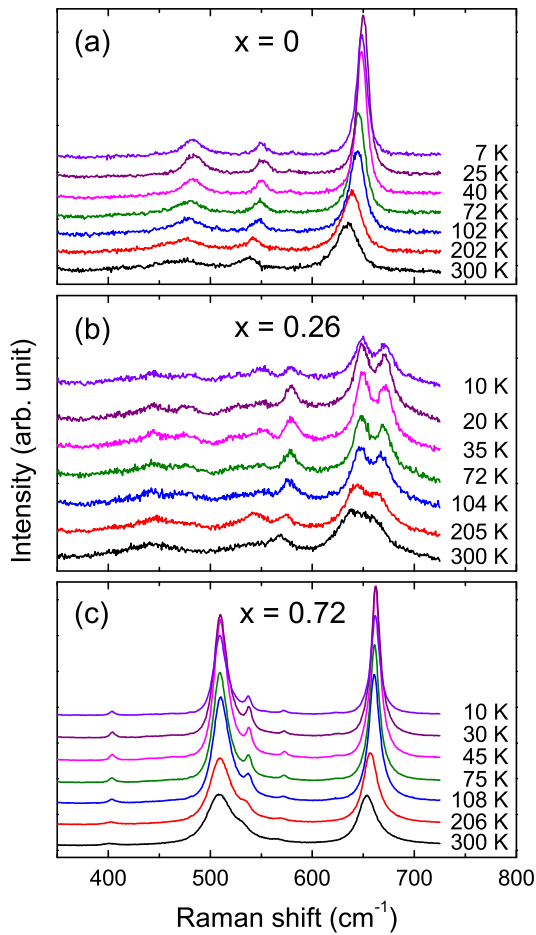}
\caption{(color online) Raman spectra at different temperatures of Ca$_3$Co$_{2-x}$Mn$_x$O$_6$ ($x =$ 0, 0.26 and 0.72).}
\end{figure}

\begin{figure}
\includegraphics[clip,width=7cm]{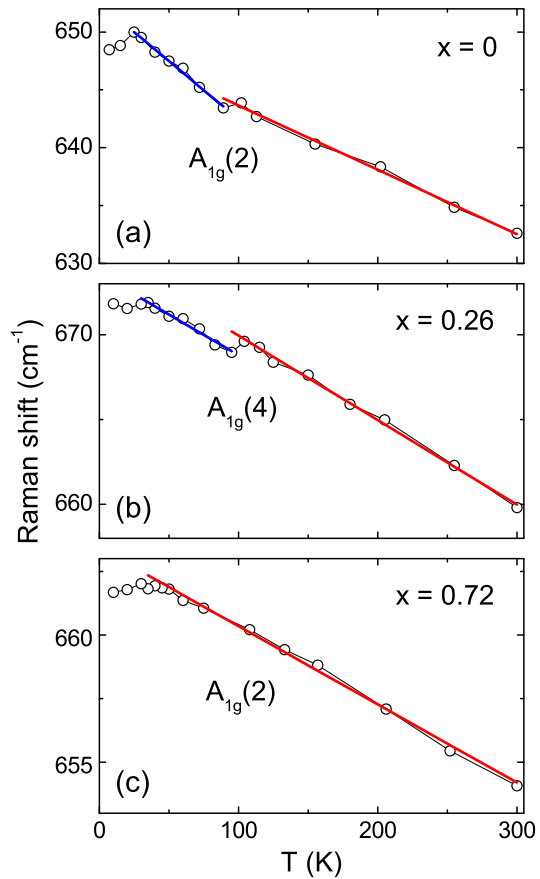}
\caption{(color online) Variations of Raman peak position as a function of temperature for Ca$_3$Co$_{2-x}$Mn$_x$O$_6$ ($x =$ 0, 0.26 and 0.72). The lines are guide to eyes for the nearly linear dependence at some temperature regimes and the anomalies show up at the deviation from these linear behaviors.}
\end{figure}

Figure 13 shows the Raman spectra at different temperatures for the $x =$ 0, 0.26 and 0.72 samples. As temperature decreases, the peaks become narrower and better resolved, due to the decrease of disorder arising from thermal motions. No new Raman modes were detected at low temperatures for all samples, indicating no structural phase transition with decreasing temperature. The spectra were also fitted by using Lorenzian function and the temperature dependencies of the $A_{1g}$(2), $A_{1g}$(4) and $A_{1g}$(2) modes are obtained and shown in Fig. 14. Each of them have anomalies with decreasing temperature. There are two sudden changes at 90 K and 25 K for the $x =$ 0 sample, as shown in Fig. 14(a). The sudden change at 90 K should be related to the anomaly observed in the extended x-ray absorption\cite{extended XAS} and M\"ossbauer measurements.\cite{Mossbauer measurements} However, the origin of this anomaly is still unclear. The existence of Co $3d$-O $2p$ hybridization and incipient magnetic order are presented in x-ray absorption and M\"ossbauer measurements, respectively. The former is understandable because the $A_{1g}$(2) mode is attributed to the vibration of Oxygen ions. The anomaly at 25 K will be discussed later. Figure 14(b) shows the variations of $A_{1g}$(4) for the $x =$ 0.26 sample. Obviously, there are also two anomalies as decreasing temperature. The lower-$T$ one is near 35 K which is coincided with $T_{FE}$. The higher-$T$ one is at about 90 K, which is consistent with the temperature of anomaly in XRD data. Since the $x =$ 0.26 sample can be treated as $x =$ 0 with low doping, it still reserve some properties of $x =$ 0 sample. For example, the Raman spectra of $x =$ 0.26 sample displays three Raman modes that are detected in the $x =$ 0 sample and shows a similar anomaly at 90 K to the $x =$ 0 sample. The above XRD results indicate that the structural anomaly at 90 K is actually a huge negative expansion effect, which is similar to the case of $x =$ 0 reported in some earlier work.\cite{extended XAS} It seems that at $T \le$ 90 K, there is overall elongation of Co-O bonds, which can be attributed to the Co $3d$-O $2p$ hybridization and the holes in the O 2$p$ bands.\cite{extended XAS} We suggest that this structure anomaly does not destroy the spatial inversion symmetry and the threefold rotational symmetry still keeps. As a result, there is probably no electric polarization below 90 K. For the $x =$ 0.72 sample, the three Raman modes disappear and two stronger $A_{1g}$ modes are detected. Figure 14(c) shows the variations of $A_{1g}$(2) for the $x =$ 0.72 sample. It just has a sudden change at 40 K, which is in agreement with the anomaly in XRD and electric polarization. Similar anomaly in Raman modes near the ferroelectric transition were also found in some other materials.\cite{DyFeCrO, YFeMnO}

\begin{figure}
\includegraphics[clip,width=6cm]{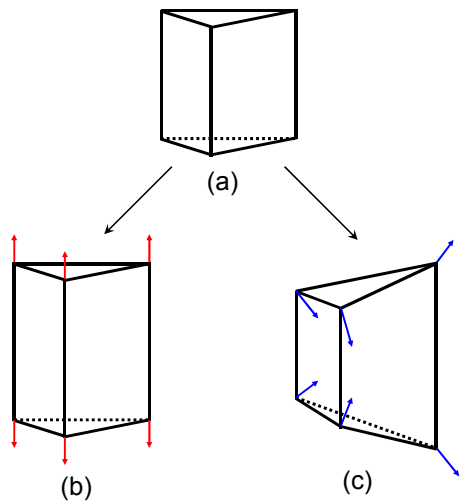}
\caption{(color online) (a) The sketch of Co$^{2+}$O$_6$ trigonal prisms.\cite{trigonal prismatic distortion} (b) The trigonal prisms just elongate along $c$ axis and do not break the threefold rotational symmetry. (c) The trigonal prisms produce a cis bond length distortion and break the threefold rotational symmetry.}
\end{figure}

\begin{figure}
\includegraphics[clip,width=8.5cm]{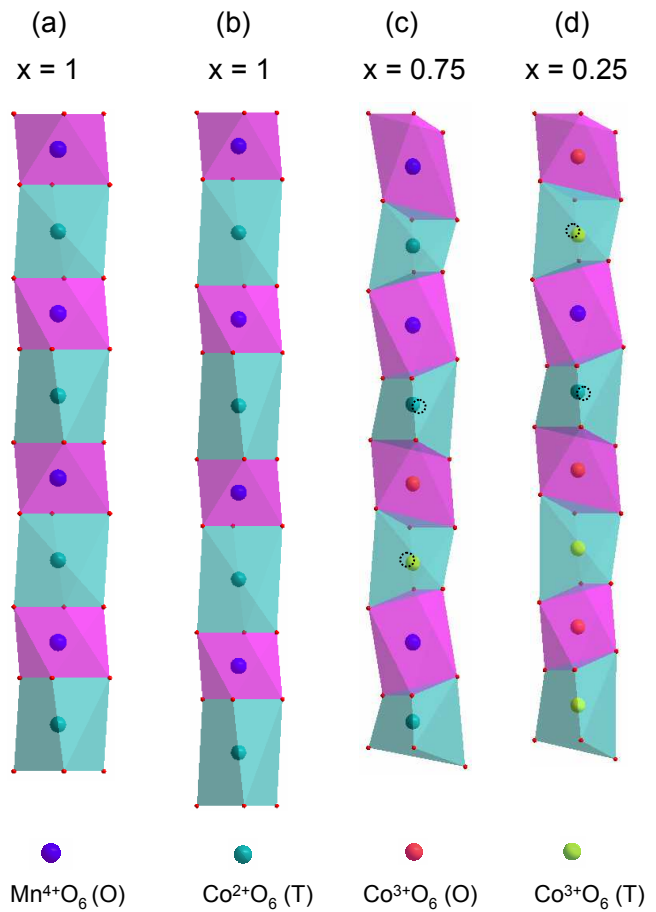}
\caption{(color online) (a) A periodic arrangement of CoMnO$_6$ chain of $x =$ 1. The Co$^{2+}$O$_6$ trigonal prisms and Mn$^{4+}$O$_6$ octahedra are alternately arranged. (b-d) Possible sketches of CoMnO$_6$ chains when the Co$^{2+}$O$_6$ trigonal prisms undergo a Jahn-Teller distortion. The Co$^{2+}$O$_6$ trigonal prisms of $x =$ 1 are elongated at the same distance on both sides. The distortions of trigonal prisms in $x =$ 0.75 and 0.25 are asymmetric. The black dashed circles are the negative-charge centers of the trigonal prisms, which are not superposed to the positive-charge centers.}
\end{figure}

Based on the XRD and Raman results, it is clear that the crystal lattices of Ca$_3$Co$_{2-x}$Mn$_x$O$_6$ ($x =$ 1, 0.72 and 0.26) have anomalous changes at low temperatures. For the $x =$ 1 sample, a clear lattice anomaly appears at 70 K. However, the electric polarization of Ca$_3$CoMnO$_6$ along the $c$ direction is known to due to the special up-up-down-down spin structure with the alternating Co$^{2+}$ and Mn$^{4+}$ ionic order at 16.5 K,\cite{CaCoMnO} instead of this lattice anomaly at 70 K. It is likely that the lattice only elongates in the $c$ direction and keeps the threefold rotational symmetry below 70 K, as shown in Fig. 15(b). In Ca$_3$CoMnO$_6$, the Co$^{2+}$O$_6$ trigonal prisms and Mn$^{4+}$O$_6$ octahedra are alternately arranged and the Co$^{2+}$O$_6$ trigonal prisms are nearly centrosymmetric surrounding the spin chains, as shown in Fig. 16(a). If the Co$^{2+}$O$_6$ trigonal prisms are elongated at the same distance on both sides, the positive-charge and negative-charge centers of the Co$^{2+}$O$_6$ trigonal prisms are still coincident, as shown in Fig. 16(b). It does not destroy the spatial inversion symmetry in this situation. Therefore, the lattice distortion at $T \le$ 70 K in Ca$_3$CoMnO$_6$ does not result in electric polarization. It is also notable that the lattice constants show an abrupt decrease at 150 K for $x =$ 1 sample. The earlier works on the  $x =$ 1 sample did not explore the electric properties around this temperature and therefore it is not clear whether this structural anomaly can induce electric polarization. In addition, it seems that at 150 K the lattice constants display quicker shrinking, which however may not necessary result in electric polarization if the local lattice symmetry is not broken.

For the $x =$ 0.72 and 0.26 samples, clear lattice anomaly can also be observed around electric polarization transition temperature ($T_{FE}$). As discussed above, the electric polarization of $x =$ 0.72 and 0.26 samples is most likely caused by a kind of Jahn-Teller distortion of Co$^{2+}$O$_6$ trigonal prisms, as shown in Figs. 15(c) and 15(d), which breaks $C_3$ rotational symmetry leading to nonzero electric polarization both along and perpendicular to the $c$ direction.\cite{trigonal prismatic distortion} Figures 16(c) and 16(d) show a periodic arrangement of CoMnO$_6$ with the particular doping of $x =$ 0.75 and 0.25, which are very close to $x =$ 0.72 and 0.26, respectively. It can be seen that in each unit cell, there is a Co$^{2+}$O$_6$ trigonal prism having different neighbors, which will lead to the asymmetric distortion of Co$^{2+}$O$_6$ trigonal prism. At these cases, the positive-charge and negative-charge centers of the trigonal prisms would not be coincident. In addition, the trigonal prisms of $x =$ 0.72 and 0.26 samples can be occupied by Co ions with different valences. As a result, a net electric polarizations can be induced with both the $ab$-plane and $c$-axis components. It can be seen that in the unit cell of both $x =$ 0.75 and 0.25 there is only a pair of Co$^{2+}$ and Co$^{3+}$ trigonal prisms, of which the electric dipoles can not be canceled with each other. This can be the reason that the electric polarizations of the $x =$ 0.72 and 0.26 samples have nearly the same value.

It should be pointed out that there is a similar anomaly of Raman shift at 25 K for the $x =$ 0 sample, as shown in Fig. 14(a). This anomaly is likely also caused by the Jahn-Teller distortion due to unevenly filled degenerate 3$d$ states of the trigonal prisms Co$^{3+}$ ions.\cite{$x =$ 0} The distortion could be similar to either the $x =$ 1 sample or $x =$ 0.72 (0.26) sample at low temperatures. If it is similar to the $x =$ 1 sample, the positive-charge and negative-charge centers of the Co$^{3+}$O$_6$ trigonal prisms are still coincident. It does not destroy the spatial inversion symmetry in this case. If it is similar to the $x =$ 0.72 (0.26) sample, the positive-charge and negative-charge centers of the trigonal prisms would not be coincident. However, all of the Co ions in $x =$ 0 sample have the valence of trivalence, there is also no net electric polarizations induced in this case.

\section{CONCLUSION}

Ca$_3$Co$_{2-x}$Mn$_x$O$_6$ single crystals with $x =$ 0.72 and 0.26 were studied by various measurements. The DC and AC magnetic susceptibility and specific heat are characteristic of the spin freezing. The electric polarization along the spin-chain direction reaches a large value of 1400 $\mu$C/m$^2$ at 8 K and the transition temperature is near 40 K and 35 K for the $x =$ 0.72 and 0.26 samples, respectively. Interestingly, an electric polarization perpendicular to the $c$ direction was also detected. However, the specific heat and magnetic susceptibility do not show anomaly at $T_{FE}$. It means that the electric polarization has no direct relationship with the magnetism. The low-temperature x-ray diffraction and the Raman spectroscopy indicate that these samples may undergo a Jahn-Teller distortion, which could be the reason of anomalous electric polarization.  Finally, it is worthy of noting that these samples display a special case of co-existence of ferroelectricity and spin glass, which would be interesting for further investigations.

\begin{acknowledgements}

This work was supported by the National Natural Science Foundation of China (Grant Nos. 11374277, U1532147, 11574286, 11404316), the National Basic Research Program of China (Grant Nos. 2015CB921201, 2016YFA0300103), the Opening Project of Wuhan National High Magnetic Field Center (Grant No. 2015KF21), and the Innovative Program of Development Foundation of Hefei Center for Physical Science and Technology. H.D.Z. thanks the support from NSF-DMR-1350002.

\end{acknowledgements}

\end{document}